 \documentclass[final,5p,times,twocolumn,authoryear]{elsarticle}

\usepackage{amsmath}
\usepackage{amsfonts}
\usepackage{amssymb}
\usepackage{graphicx}
\usepackage[titletoc]{appendix}
\usepackage{color}
\usepackage{hyperref}
\usepackage{cleveref}
\usepackage[rightcaption]{sidecap}
\usepackage{subfigure}
\usepackage{comment}
\usepackage{soul}
\usepackage{cancel}
\usepackage{dcolumn}
\usepackage{multicol}

\usepackage{array}
\usepackage{ctable}
\usepackage{multirow}
\usepackage{siunitx}
\usepackage{longtable}
\usepackage{tabularx}
\usepackage{booktabs}
\usepackage{float}

\graphicspath{{Graphics/}}

\def\be{\begin{equation}}
\def\ee{\end{equation}}
\def\bea{\begin{eqnarray}}
\def\eea{\end{eqnarray}}

\def\prl{Phys. Rev. Lett.}

\def\prd{Phys. Rev. D}
\def\mnras{MNRAS}

\def\apj{ApJ}
\def\apjl{ApJ Lett.}
\def\apjs{ApJ Suppl. Ser.}

\def\aap{A\&A}

\def\jcap{JCAP}

\def\apss{Astrophysics and Space Science}
\def\jcp{J. Chem. Phys.}

\definecolor{vividviolet}{rgb}{0.62, 0.0, 1.0}
\definecolor{amaranth}{rgb}{0.9, 0.17, 0.31}
\definecolor{palatinateblue}{rgb}{0.15, 0.23, 0.89}
\definecolor{brightpink}{rgb}{1.0, 0.0, 0.5}
\definecolor{cornflowerblue}{rgb}{0.39, 0.58, 0.93}
\definecolor{deepcarminepink}{rgb}{0.94, 0.19, 0.22}
\definecolor{radicalred}{rgb}{1.0, 0.21, 0.37}

\hypersetup{ linktoc=all,
    colorlinks, linkcolor={palatinateblue},
    citecolor={brightpink}, urlcolor={amaranth}
}

\DeclareUnicodeCharacter{2212}{\ensuremath{-}}
\DeclareUnicodeCharacter{2243}{\ensuremath{\simeq}}

\journal{High Energy Astrophysics}

\begin{document}

\begin{frontmatter}



\title{Investigating the cosmic distance duality relation with gamma-ray bursts}


\author[1,2]{Anna Chiara Alfano}
\ead{a.alfano@ssmeridionale.it}
\author[3]{Carlo Cafaro}
\ead{ccafaro@albany.edu}
\author[1,2,4,5]{Salvatore Capozziello}
\ead{capozziello@na.infn.it}
\author[3,6,7,8,9]{Orlando Luongo}
\ead{orlando.luongo@unicam.it}
\author[6,9,10]{Marco Muccino}
\ead{marco.muccino@lnf.infn.it}

\affiliation[1]{organization={Scuola Superiore Meridionale},
            addressline={Largo S. Marcellino 10},
            city={Napoli},
            postcode={80138},
            country={Italy}}

\affiliation[2]{organization={Istituto Nazionale di Fisica Nucleare (INFN), Sezione di Napoli Complesso Universitario Monte S. Angelo},
            addressline={Via Cinthia 9 Edificio G},
            city={Napoli},
            postcode={80126},
            country={Italy}}

\affiliation[3]{organization={University at Albany-SUNY},
city={Albany, NY},
postcode={12222},
country={USA}}

\affiliation[4]{organization={Dipartimento di Fisica “E. Pancini”, Università di Napoli “Federico II”, Complesso Universitario di Monte Sant’Angelo, Edificio G},
addressline={Via Cinthia},
postalcode={I-80126},
city={Napoli},
country={Italy.}}

\affiliation[5]{organization={Research Center of Astrophysics and Cosmology, Khazar University},
city={Baku},
postalcode={AZ1096},
addressline={41 Mehseti Street},
country={Azerbaijan}.}

\affiliation[6]{organization={Università di Camerino, Divisione di Fisica},
            addressline={Via Madonna delle carceri 9},
            city={Camerino},
            postcode={62032},
            country={Italy}}

\affiliation[7]{organization={INFN, Sezione di Perugia},
            city={Perugia},
            postcode={06123},
            country={Italy}}

\affiliation[8]{organization={INAF - Osservatorio Astronomico di Brera},
            city={Milano},
            country={Italy}}

\affiliation[9]{organization={Al-Farabi Kazakh National University},
            addressline={Al-Farabi av. 71},
            city={Almaty},
            postcode={050040},
            country={Kazakhstan}}

\affiliation[10]{organization={ICRANet},
            addressline={P.zza della Repubblica 10},
            city={Pescara},
            postcode={65122},
            country={Italy}}

\begin{abstract}
Deviations from the so-called {\it cosmic distance duality relation} may result from systematic errors in distance measurements or, more interestingly, hint at new physics. Further, it can also be related to the Hubble constant tension between early and local measurements of $H_0$. Based on this, we test validity of this relation through a model-independent parameterization of the Hubble rate via the well-estabilished B\'ezier polynomials approach. We seek for possible departures from the relation considering three parametrizations, i) a power-law correction, ii) a logarithmic correction and iii) a Pad\'e series $P_{n,m}(z)$ of order (1;2) with $n=1$ being the order of the numerator while $m=2$ is the order of the denominator. Then, assuming a flat scenario, we test them through Monte Carlo -- Markov chain analyses that combine low- and intermediate/high-$z$ data sets, such as observational Hubble data, the Pantheon catalog of type Ia supernovae, galaxy clusters, the second data release from the DESI Collaboration and gamma-ray bursts. In particular, we distinguish between \emph{Analysis A} and \emph{Analysis C}, depending whether the prompt emission $E_{iso}-E_p$ or the prompt-afterglow $L_0-E_p-T$ gamma-ray burst correlations, respectively, is fit together with the other probes previously described. Our results seem to point towards a \emph{no violation} of the cosmic distance duality relation and a preference towards Planck's value of $H_0$.
\end{abstract}




\end{frontmatter}




\section{Introduction}

The cosmic distance duality (CDD) relation, also known under the name of Etherington's reciprocity law \citep{etherington1933lx}, relates different definitions of cosmic distances. In its more discussed form, the relation focuses on the luminosity distance $D_L$ and the angular diameter distance $D_A$ as
\begin{equation}\label{CDD}
    \frac{D_L}{D_A} = a^{-2} = (1+z)^2,
\end{equation}
where $a(t)\equiv(1+z)^{-1}$ is the scale factor of the universe and $z$ is the cosmic redshift. This relation is valid for any metric theory of gravity  implying that photons travel along null geodesics and that their number between the source and the observer is conserved \citep{ellis2007definition, 2015PhRvD..92l3539L}. Thus, if a violation of Eq.~\eqref{CDD} would subsist, it could point to the presence of systematic errors or exotic physics \citep{ellis2007definition, 2004PhRvD..69j1305B}.

Moreover, deviations from the CDD relation can also be linked to a thorny problem affecting the $\Lambda$CDM model, namely the $H_0$ tension\footnote{Measurements of $h_0=H_0/(100~{\rm km/s/Mpc})$ at early-times from the \citet{2020A&A...641A...6P}, i.e. $h^P_0=0.674\pm 0.005$, and at late-times using type Ia supernovae (SNe~Ia) calibrated through Cepheids \citep{2022ApJ...934L...7R}, i.e. $h^S_0=0.730\pm 0.010$, show a $\sim5\sigma$ discrepancy. Many efforts have been put into alleviate or solving this tension, for an extensive review on this topic see Ref.~\cite{2021CQGra..38o3001D}
and the more recent \cite{DiValentino:2025sru}. See also \cite{Capozziello:2023ewq,Capozziello:2024stm} for alternative explanations and possible solutions.}
\citep{2021APh...13102605D}  that can be seen as a \emph{cosmic calibration tension} arising when luminosity and angular diameter distance measurements are compared, i.e. when the CDD relation is taken into consideration \citep{2024arXiv240718292P, Teixeira:2025czm}.

In order to investigate possible deviations from Eq.~\eqref{CDD} the basic procedure is to consider that a departure from the CDD relation is in the form
\begin{equation}\label{cddviol}
    \frac{D_L}{D_A(1+z)^2}=\eta(z),
\end{equation}
where $\eta(z)$ can assume various forms \citep{2011A&A...528L..14H, 2012IJMPD..2150008H, 2012PhRvD..85l3510C, 2014JPhCS.484a2035J, 2015JCAP...10..061S, 2017JCAP...09..039H, 2020PhRvD.102f3513D, 2022EPJC...82..115H, 2024EPJC...84..702W, 2025JCAP...01..088J, Teixeira:2025czm} to test and assess the robustness of the CDD relation. For one or two parameter representations \citep{2012PhRvD..85l3510C, 2014JPhCS.484a2035J, 2020PhRvD.102f3513D} and parametrizations aiming at exploring deviations at low-$z$ regimes, inspired by the parametrizations of the dark energy barotropic factor, see e.g. Ref.~\cite{2025JCAP...01..088J}, or by taking into consideration possible departures from the \emph{cosmic transparency} \citep{2009JCAP...06..012A}.
Besides exploring different parametrizations of $\eta(z)$, many authors investigated both model-dependent and model-independent tests of a possible violation of Eq.~\eqref{cddviol}, i.e. when $\eta(z)\neq 1$ \citep{2004PhRvD..70h3533U, 2006IJMPD..15..759D, 2016PDU....13..139L, 2012ApJ...745...98M, 2017JCAP...07..010R, 2018ApJ...866...31R, 2021MNRAS.504.3938M, 2022ApJ...939..115X, 2024PhLB..85839027F, 2025JCAP...01..088J}.

In this work, we investigate three parametrizations of $\eta(z)$: i) the power-law (PL) parametrization that also takes into account departures from the cosmic transparency, ii) the LOG model aiming to alleviate the convergence problem arising from using a Taylor expansion of i), and iii) the Pad\'e polynomials addressing even further the convergence problem from ii) \citep{2014PhRvD..89j3506G, 2014PhRvD..90d3531A} and providing the strongest constraints on the parameters.

Furthermore, in the literature, Eq.~\eqref{cddviol} has been mostly tested with low-$z$ data sets, albeit intermediate/high-$z$ sources would serve as powerful tools to investigate even further possible CDD violations, see e.g. Ref \cite{2017JCAP...09..039H}.
To this aim, we hereby expand this test including gamma-ray bursts (GRBs) in our analysis. These transients can be found at redshifts as high as $z\simeq 9$ and satisfy various correlations involving prompt and prompt-afterglow emission observables \citep{2002A&A...390...81A, 2006MNRAS.372..233A, 2015A&A...582A.115I, 2004ApJ...609..935Y, 2004ApJ...616..331G, 2008MNRAS.391L..79D,Dainotti:2024wut,
Bargiacchi:2024srw, Dainotti:2023zep, Dainotti:2022ked,Cao:2022yvi}.

Among the plethora of correlations we focus our attention on the $E_p-E_{iso}$ (or \emph{Amati}) correlation which falls under the group of prompt emission correlations and the $L_0-E_p-T$ (or \emph{Combo}) correlation which falls under the group of prompt-afterglow emission correlations.

We investigate only these two correlations because:
\begin{itemize}
    \item [-] prompt and prompt-afterglow correlations may furnish slightly different results for the CDD relation,
    \item [-] the $E_p-E_{iso}$ correlation is the most used prompt emission correlation in the literature \citep{2010A&A...519A..73C, 2011MNRAS.415.3580D, 2021JCAP...09..042K, 2021A&A...651L...8D, 2023MNRAS.523.4938M, 2024JCAP...12..055A, 2024arXiv241220424L, 2024arXiv241218493L}, and
    \item [-] the $L_0-E_p-T$ correlation is one of the most promising among the prompt-afterglow emission ones.
\end{itemize}

Even though GRBs would serve as high-redshift probes, they suffer from a \emph{circularity} problem jeopardizing their use as distance indicators since, unlike SNe~Ia, the lack of GRBs at low-$z$ does not permit to anchor them to primary distance indicators.
For the $E_p-E_{iso}$ and the $L_0-E_p-T$ correlations this issue rises because the radiated isotropic energy $E_{iso}$ in the prompt emission and the X-ray afterglow plateau luminosity $L_0$, respectively, depend on an underlying cosmological model \citep{2021Galax...9...77L}. To circumvent this problem, various approaches were proposed throughout the literature \citep{2015GReGr..47..141L, 2017A&A...598A.112D, 2017A&A...598A.113D, 2023MNRAS.521.4406L, 2025Ap&SS.370...10Z}, among which the Hubble rate $H(z)$ reconstruction based on B\'ezier polynomials seems one of the most promising one \citep{2019MNRAS.486L..46A, 2020A&A...641A.174L, 2021MNRAS.503.4581L, 2021MNRAS.501.3515M, 2023MNRAS.518.2247L, 2023MNRAS.523.4938M, 2024JHEAp..42..178A}.

With this in mind, we here apply the B\'ezier interpolation to the observational Hubble rate data (OHD) and propagate this model-independent reconstruction to the other catalogs employed in this work, including GRB data from the $E_p-E_{iso}$ and the $L_0-E_p-T$ correlations.
Thus, to constrain the CDD parameters, as well as the B\'ezier coefficients and the GRB correlation parameters, we run two Monte-Carlo Markov chain (MCMC) analyses based on the Metropolis-Hastings algorithm \citep{1953JChPh..21.1087M, 1970Bimka..57...97H}:the Pantheon catalog of SNe~Ia \citep{2018ApJ...859..101S}, the catalog of angular diameter distances derived considering the Sunyaev-Zeldovich (SZ) effect \citep{2005ApJ...625..108D}, and the second data release (DR2) of the baryonic acoustic oscillations (BAO) data from the \citet{2025arXiv250314738D} are combined with GRBs from the $E_p-E_{iso}$ data set in \emph{Analysis A} (as in \emph{Amati}), and with the $L_0-E_p-T$ data set in \emph{Analysis C} (as in \emph{Combo}).

From our results, no violation of the CDD relation is observed for both \emph{Analysis A} and \emph{Analysis C}. Regarding the (reduced) Hubble constant $h_0$ we find that, for both \emph{Analyses A} and \emph{C} and for all three parametrizations of $\eta(z)$, the constraints agree with the value $h^P_0=0.674\pm 0.005$  from the \citet{2020A&A...641A...6P} at $1$-$\sigma$. We also compared our $h_0$ with the one inferred by using SNe Ia, i.e. $h_0^S=0.730\pm 0.010$ \citep{2022ApJ...934L...7R} finding no compatibility between $h_0^S$ and our inferred values.

The paper is organized as follows. In Sect.~\ref{sec1} we discuss the CDD relation and present the utilized parametrizations of $\eta(z)$. Then, Sect.~\ref{sec2} deals with the model-independent approach of B\'ezier interpolation with extensive discussions on each probe used for our computations. Afterwards, Sect.~\ref{sec3} describes the two GRB correlations adopted in this work while  Sect.~\ref{sec4} discusses the outcomes of our MCMC analyses. Finally, Sect.~\ref{sec5} focuses on the conclusions emerging from this work.

\section{Cosmic distance duality parametrizations}\label{sec1}

As already discussed, possible departures from the CDD relation would hint at the investigation of new physics.
As a matter of fact, the relation is valid for any metric theories of gravity: light travels along null geodesics and the photon number is conserved \citep{ellis2007definition, 2015PhRvD..92l3539L}. In this regard, a violation of the CDD relation could arise from photon interactions with light particles such as chameleons or axion-like particles \citep{2002PhRvL..88p1302C,2004ApJ...607..661B, 2008PhRvD..77d3009B, 2010JCAP...10..024A,Addazi:2024mii}, photon-graviton conversion induced by a primordial magnetic field \citep{1995PhRvL..74..634C, 1996PhRvD..54.4757C,Addazi:2024osi,Addazi:2024kbq,Capozziello:2022dle} also considered in universes with extra dimensions such as Kaluza-Klein models \citep{2000PhRvD..62f3507D} and so on.
Hence, to investigate possible departures from the relation many parametrizations of $\eta(z)$ were proposed throughout the literature \citep{2011A&A...528L..14H, 2012IJMPD..2150008H, 2012PhRvD..85l3510C, 2014JPhCS.484a2035J, 2015JCAP...10..061S, 2017JCAP...09..039H, 2020PhRvD.102f3513D, 2022EPJC...82..115H, 2024EPJC...84..702W, 2025JCAP...01..088J, Teixeira:2025czm}.

In this work we consider as first parametrization
\begin{equation}\label{stepintermedio}
    \eta(z)\equiv (1+z)^{\eta_0},
\end{equation}

where $\eta_0$ is a parameter that quantifies possible deviations from the CDD relation. Further, Eq. \eqref{stepintermedio} is also used to investigate possible violation of the conservation of the number of photons as it is linked to deviations from the \emph{cosmic transparency} \citep{2009JCAP...06..012A}. Moreover, departures from $\eta_0=0$ in this scenario are also investigated in terms of the temperature of the cosmic microwave background (CMB) radiation as a direct consequence of the violation of the photon number \citep{2020A&A...644A..80M, 2018PhRvD..97b3538H, 2020JCAP...08..009A}.

Afterwards, we rewrite Eq.~\eqref{stepintermedio} in the form of the so-called LOG model considering that $|\eta_0|\ll 1$
\begin{equation}
    \eta(z)=\exp\{\eta_0\ln(1+z)\}\simeq 1+\eta_0\ln(1+z).
\end{equation}
where the correction adopted using the natural logarithm implies a weaker evolution in terms of $z$ in both low- and high-redshift regimes. However, this correction does not fully address and heal the convergence problem. Therefore, we resort to the so-called Pad\'e polynomials, widely used in the literature as a cosmology-independent technique to write the luminosity distance \citep{2014PhRvD..90d3531A, 2014JCAP...01..045W, 2017ApJ...843...65R, 2018JCAP...05..008C, 2019arXiv190305860L, 2020MNRAS.494.2576C, Petreca:2023nhy, Benetti:2019gmo, Capozziello:2018jya}. First introduced in Ref.~\cite{2014PhRvD..89j3506G}, this approach aims at healing systematics raising from the truncation of the Taylor series providing in this way better constraints on the cosmographic parameters\footnote{Other methods proposed to address the convergence problem are reparametrizations of $z$ via auxiliary variables \citep{2007CQGra..24.5985C, 2012PhRvD..86l3516A}. However, these methods fail to converge at redshift larger than $z=1$, where the majority of probes are found.}.

With this in mind, we write Eq.~\eqref{stepintermedio} by means of a Pad\'e rational polynomial of order $(1;2)$
\begin{equation}\label{P12}
    {P}_{1,2}(z)=\eta^{(1;2)}(z)\simeq \frac{6 + 2 z (2 + \eta_0)}{6 + 4 z (1 - \eta_0) - z^2 (1 - \eta_0) \eta_0}.
\end{equation}

The standard CDD relation is resumed as soon as  $\eta_0=0$ for the three parametrizations described above.

\section{Model-independent B\'ezier interpolation}\label{sec2}

In this work, we do not assume any background cosmology, but rather we approximate the Hubble rate in a model-independent approach by means of a B\'ezier curve of order $n=2$, which is the only order providing a non-linear monotonic behavior \citep{2025A&A...693A.187L,2019MNRAS.486L..46A,2020A&A...641A.174L,2021MNRAS.503.4581L,2024arXiv241220424L, 2024A&A...686A..30A},
\begin{equation}\label{bezier}
H_2(z)=\frac{g_\alpha}{z^2_M}[\alpha_0(z_M-z)^2+2\alpha_1z(z_M-z)+\alpha_2z^2],
\end{equation}
where $g_\alpha=100 \ \text{km/s/Mpc}$ and $z_M$ is the maximum redshift of the OHD catalog. Further, the coefficients appearing in Eq. \eqref{bezier}, i.e. $\alpha_0, \ \alpha_1$ and $\alpha_2$ are the  B\'ezier coefficients.

Our recipe is to apply this methodology to the data used in this work and maximizing the total ln-likelihood to find the best fit parameters. Specifically, a generic likelihood $\mathcal{L}$ can be written as $\mathcal{L}\propto\text{exp}(-\chi^2/2)$ where the $\chi^2$ varies depending on whether the measurements considered are correlated or not
\begin{equation}\label{chi2}
    \chi^2 = \begin{cases}
        \sum_i \left[X_i - X(z_i)\right]^2\sigma_i^{-2}, & \text{uncorrelated}, \\[1ex]
        \sum_i \Delta_i^{\text{T}} \mathbf{C}^{-1} \Delta_i, & \text{correlated},
    \end{cases}
\end{equation}
where $\Delta_i=X_i-X(z_i)$, $X_i$ are the data points with $\sigma_i$ errors, while $X(z_i)$ are the theoretical counterparts.

Below we proceed to describe the adopted data and their ln-likelihoods, i.e. $\ln\mathcal{L}\propto-\chi^2/2$.

\begin{enumerate}
    \item [-] {\bf OHD.} The most updated catalog consists of $N_O=34$ data points (see Tab.~\ref{tab:OHD}). These model-independent measurements of the Hubble rate $H(z)$ are derived by considering as cosmic chronometers red galaxies which evolve passively and measuring the difference in age of close pairs of such galaxies at different redshifts $z$ through the relation $H(z)=-(1+z)^{-1}(dz/dt)$ \citep{2002ApJ...573...37J}.

    The OHD full covariance matrix $\text{C}^{\text{tot}}_{ij}=\text{C}_{ij}^{\text{stat}}+\text{C}_{ij}^{\text{sys}}$ encapsulates both statistical errors through the diagonal matrix  $\text{C}_{ij}^{\text{stat}}$ and systematic contributions $\text{C}_{ij}^{\text{sys}}$ due to several effects arising from \citep{2023arXiv230709501M}:

    \begin{itemize}
        \item [-] the metallicity of the sample, given by $\text{C}_{ij}^{\text{met}}$,
        \item [-] a residual contamination from a younger component defined as $\text{C}_{ij}^{\text{young}}$,
        \item [-] the assumed model for stellar population synthesis (SPS) $\text{C}_{ij}^{\text{SPS}}$, stellar libraries $\text{C}_{ij}^{\text{st,lib}}$, the initial mass function (IMF) $\text{C}_{ij}^{\text{IMF}}$ and the star formation history (SFH) $\text{C}_{ij}^{\text{SFH}}$.
    \end{itemize}

    Among the above  contributions, $\text{C}_{ij}^{\text{met}}$, $\text{C}_{ij}^{\text{young}}$ and $\text{C}_{ij}^{\text{SFH}}$ are diagonal matrices and their contributions are already included in the errors \citep{2022LRR....25....6M, 2023arXiv230709501M}; the other contributions introduce non-diagonal terms. The contributions derived from the IMF are $\lesssim0.5\%$ while SPS and stellar libraries are those who primarily contributes to the errors in the sample. However, in \citet{2020ApJ...898...82M} the calculated covariance matrix only took into account the 15 OHD data points available at that time and since the catalog has been updated, this covariance matrix does not represent the entire sample \citep{2025arXiv250211443N}.
    However, as done in \citet{2023MNRAS.523.4938M}, these systematics have been duly propagated throughout the whole data set, thus the total errors are obtained adding in quadrature both statistical and systematic errors listed in Tab.~\ref{tab:OHD}.

    Fitting OHD with the B\'ezier curve in Eq.~\eqref{bezier} provides a model-independent estimate of $h_0$ since at $z=0$ we have $\alpha_0\equiv h_0$ \citep{2021MNRAS.503.4581L}.

    In this way, the corresponding ln-likelihood for this catalog is
    \begin{equation}
    \ln\mathcal{L}_O\propto-\frac{1}{2}\sum^{N_O}_{i=1}\left[\frac{H_i-H_2(z_i)}{\sigma_{H_i}}\right]^2,
    \end{equation}
    where $H_i$ are the OHD data points with errors $\sigma_{H_i}$ while $H_2(z_i)$ is given by Eq.~\eqref{bezier}.

   \begin{table}[t]
    \centering
    \setlength{\tabcolsep}{.5em}
    \renewcommand{\arraystretch}{1.}
   \begin{tabular}{lcc}
   \hline
    $z$     &$H(z)$ &  Refs. \\
    \hline
    0.07  & $69.0  \pm 19.6\pm 12.4$ & \cite{Zhang2014} \\
    0.09    & $69.0  \pm 12.0\pm 11.4$  & \cite{Jimenez2002} \\
    0.12    & $68.6  \pm 26.2\pm 11.4$  & \cite{Zhang2014} \\
    0.17    & $83.0  \pm 8.0\pm 13.1$   & \cite{Simon2005} \\
    0.1791   & $75.0  \pm 3.8\pm 0.5$   & \cite{Moresco2012} \\
    0.1993   & $75.0  \pm 4.9\pm 0.6$   & \cite{Moresco2012} \\
    0.20    & $72.9  \pm 29.6\pm 11.5$  & \cite{Zhang2014} \\
    0.27    & $77.0  \pm 14.0\pm 12.1$  & \cite{Simon2005} \\
    0.28    & $88.8  \pm 36.6\pm 13.2$  & \cite{Zhang2014} \\
    0.3519   & $83.0  \pm 13.0\pm 4.8$  & \cite{Moresco2016} \\
    0.3802  & $83.0  \pm 4.3\pm 12.9$  & \cite{Moresco2016} \\
    0.4     & $95.0  \pm 17.0\pm 12.7$  & \cite{Simon2005} \\
    0.4004  & $77.0  \pm 2.1\pm 10.0$  & \cite{Moresco2016} \\
    0.4247  & $87.1  \pm 2.4\pm 11.0$  & \cite{Moresco2016} \\
    0.4497  & $92.8  \pm 4.5\pm 12.1$  & \cite{Moresco2016} \\
    0.47    & $89.0\pm 23.0\pm 44.0$     & \cite{2017MNRAS.467.3239R}\\
    0.4783  & $80.9  \pm 2.1\pm 8.8$   & \cite{Moresco2016} \\
    0.48    & $97.0  \pm 62.0\pm 12.7$  & \cite{Stern2010} \\
    0.5929   & $104.0 \pm 11.6\pm 4.5$  & \cite{Moresco2012} \\
    0.6797    & $92.0  \pm 6.4\pm 4.3$   & \cite{Moresco2012} \\
    0.75    & $98.8\pm33.6$     & \cite{2022ApJ...928L...4B}\\
    0.7812   & $105.0 \pm 9.4\pm 6.1$  & \cite{Moresco2012} \\
    0.80    & $113.1\pm15.1\pm 20.2$    & \cite{2023ApJS..265...48J}\\
    0.8754   & $125.0 \pm 15.3\pm 6.0$  & \cite{Moresco2012} \\
    0.88    & $90.0  \pm 40.0\pm 10.1$  & \cite{Stern2010} \\
    0.9     & $117.0 \pm 23.0\pm 13.1$  & \cite{Simon2005} \\
    1.037   & $154.0 \pm 13.6\pm 14.9$  & \cite{Moresco2012} \\
     1.26    & $135.0\pm 65.0$          & \cite{Tomasetti2023}\\
    1.3     & $168.0 \pm 17.0\pm 14.0$  & \cite{Simon2005} \\
    1.363   & $160.0 \pm 33.6$  & \cite{Moresco2015} \\
    1.43    & $177.0 \pm 18.0\pm 14.8$  & \cite{Simon2005} \\
    1.53    & $140.0 \pm 14.0\pm 11.7$  & \cite{Simon2005} \\
    1.75    & $202.0 \pm 40.0\pm 16.9$  & \cite{Simon2005} \\
    1.965   & $186.5 \pm 50.4$  & \cite{Moresco2015} \\
\hline
\end{tabular}
\caption{The updated sample of OHD where the first column respectively displays the redshift, the values of $H(z)$ together with the statistical and systematic errors are in the second column, and the references from which these values were taken appear in the third column.}
\label{tab:OHD}
\end{table}

\item [-] {\bf Galaxy clusters data.} Direct measurements of the angular diameter distances $D_A(z)$ are found combining microwave observations of the CMB spectrum distortion, caused by the inverse Compton scattering between CMB photons and the electrons in the medium of galaxy clusters, with the X-ray emission from intra-cluster electrons.
This phenomenon is the well-known Sunyaev--Zeldovich (SZ) effect \citep{1970CoASP...2...66S, 1972CoASP...4..173S}.

Using the SZ effect, $N_{GC} = 25$ angular diameter distance measurements were found for the galaxy clusters listed in Tab.~\ref{tab:SZ}  \citep{2005ApJ...625..108D}.
However, these distances are biased by the validity of the CDD relation \citep{2004PhRvD..70h3533U} leading to consider the inferred angular diameter distances in the following form \citep{2011A&A...528L..14H, 2012IJMPD..2150008H}
\begin{equation}\label{damod}
    D^{GC}_A(z) = \eta^2(z)D_A(z),
\end{equation}
where $\eta(z)$ depends upon the parametrization. The angular diameter distance $D_A(z)$, within our model-independent approach, takes the form
\begin{equation}\label{da}
   D_2(z) = \frac{c(100\alpha_0)^{-1}}{(1+z)\sqrt{|\Omega_k|}}S_k\left[\int^z_0 \sqrt{|\Omega_k|}\frac{dz^\prime}{E_2(z^\prime)}\right],
\end{equation}
where $c$ is the speed of light,  $E_2(z)\equiv H_2(z)/(100\alpha_0)$ is the normalized Hubble rate, and $S_k(y)$ varies according to the sign of $\Omega_k$
\begin{equation}
    S_k(y)= \begin{cases}
    {\sinh(y)}, & \text{for} \quad \Omega_k > 0, \\
    {\sin(y)}, & \text{for} \quad \Omega_k < 0, \\
    y, & \text{for} \quad \Omega_k = 0.
    \end{cases}
\end{equation}
For our analysis, we limit ourselves to consider the flat case, i.e. $\Omega_k=0$, so that $S_k(y)\equiv y$.

\begin{table}[t]
\centering
\setlength{\tabcolsep}{3.em}
\renewcommand{\arraystretch}{1.}
\begin{tabular}{lc}
\hline
$z$     & $D_A(z)$ \\
\hline
$0.023$	& $103\pm42$\\
$0.058$	& $242\pm61$\\
$0.072$ & $165\pm45$\\
$0.074$ & $369\pm62$\\
$0.084$	& $749\pm385$\\
$0.088$	& $448\pm185$\\
$0.091$	& $335\pm70$\\
$0.142$	& $478\pm126$\\
$0.176$	& $809\pm263$\\
$0.182$	& $451\pm189$\\
$0.183$ & $604\pm84$\\
$0.202$	& $387\pm141$\\
$0.202$	& $806\pm163$\\
$0.217$	& $1465\pm407$\\
$0.224$	& $1118\pm283$\\
$0.252$	& $946\pm131$\\
$0.282$	& $1099\pm308$\\
$0.288$	& $934\pm331$\\
$0.322$	& $885\pm207$\\
$0.327$	& $697\pm183$\\
$0.375$	& $1231\pm441$\\
$0.451$	& $1166\pm262$\\
$0.541$	& $1635\pm391$\\
$0.550$	& $1073\pm238$\\
$0.784$	& $2479\pm1023$\\
\hline
\end{tabular}
\caption{The sample of $25$ galaxy clusters. The first column portrays the redshifts, while the second column shows the angular diameter distances with attached errors \citep{2005ApJ...625..108D}.}
\label{tab:SZ}
\end{table}

Thus, the ln-likelihood for the galaxy cluster data points $D_{Ai}$ with errors $\sigma_{D_{Ai}}$ \citep{2005ApJ...625..108D} is given by
\begin{equation}
    \ln\mathcal{L}_{GC} \propto -\frac{1}{2}\sum^{N_{GC}}_{i = 1}\left[\frac{D_{Ai}-\eta^2(z_i)D_2(z_i)}{\sigma_{D_{Ai}}}\right]^2,
\end{equation}
where the distances $D_2(z_i)$ are defined by Eq.~\eqref{da}.

\item [-] {\bf SNe~Ia.} One of the most updated samples of SNe~Ia is the Pantheon catalog \citep{2018ApJ...859..101S}, consisting of 1048 sources with associated distance moduli defined as
\begin{equation}
    \mu_{SN}=m_B+(\alpha\chi_1-\beta\mathcal{C}+\Delta_M+\Delta_B-\mathcal{M}),
\end{equation}
where $m_B$ is the $B$-band apparent magnitude, $\alpha$ and $\beta$ define the luminosity-stretch and the luminosity-color coefficients, respectively. These coefficients multiply the factors $\chi_1$ and $\mathcal{C}$, respectively. On the other hand, $\Delta_M$ and $\Delta_B$ are distance corrections which are based on the mass of the galaxy hosting the SNe Ia and biases, respectively.

The ln-likelihood for this sample is obtained by marginalizing over the $B$-band absolute magnitude $\mathcal{M}$ \citep{2011ApJS..192....1C}
\begin{equation}
    \ln\mathcal{L}_{SN}=-\frac{1}{2}\left(a-\frac{b^2}{e}+\ln\frac{e}{2\pi}\right),
\end{equation}
where $a\equiv\Delta\mu_S^\text{T}\textbf{C$_\textbf{S}$}^{-1}\Delta\mu_S$, $b\equiv\Delta\mu_S^\text{T}\textbf{C$_\textbf{S}$}^{-1}{\bf 1}$ and $e\equiv {\bf 1}^{\text{T}}\textbf{C$_\textbf{S}$}^{-1}{\bf 1}$ where $\textbf{C$_\textbf{S}$}$ is the covariance matrix of the catalog while $\Delta\mu_S=\mu_S-\mu_S^\text{th}$ with $\mu_S^\text{th}$ being the theoretical distance moduli which in our case, considering Eq. \eqref{cddviol} becomes
\begin{equation}
    \mu_S^\text{th}=25+5\log[\eta(z)(1+z)^2D_2(z)],
\end{equation}
where $D_2(z)$ is, as always, defined in Eq. \eqref{da}.

\item [-] {\bf DR2 DESI-BAO.} The DR2 DESI consists of $N_D = 19$ values of the transverse comoving distance, the Hubble rate distance and the angle-averaged distance, i.e. $12$ data from the first release with reduced errors and $7$ additional data points \citep{2025arXiv250314738D}. In our analyses we consider only the $N_D=13$ distance values displayed in Tab.~\ref{tab:DESIBAO}, for which a covariance matrix is available.

To break the $r_d-h_0$ degeneracy \citep{2024arXiv240403002D} and constrain $h_0$, we fix the sound horizon at the baryon drag epoch to $r_d = (147.09\pm 0.26)\ \text{Mpc}$ \citep{2020A&A...641A...6P}.
Considering the violation of the CDD relation the three distances are written as
\begin{subequations}\label{distances}
    \begin{align}
        \frac{D_M(z)}{r_d} &= \frac{\eta(z)D_2(z)(1+z)}{r_d},\label{dm}\\
        \frac{D_H(z)}{r_d}& = \frac{c}{r_dH_2(z)},\\
        \frac{D_V(z)}{r_d} &= \frac{1}{r_d}\left[\frac{zc}{H_2(z)}\right]^{\frac{1}{3}}\left[\eta(z)D_2(z)(1+z)\right]^{\frac{2}{3}}.    \end{align}
\end{subequations}
In this case, the ln-likelihood is
\begin{equation}
    \ln\mathcal{L}_D\propto -\frac{1}{2}\sum^{N_D}_{i=1}\Delta Y_i^\text{T}\textbf{C}_\textbf{B}^{-1}\Delta Y_i,
\end{equation}
with $\Delta Y_i=Y_i-Y(z_i)$ where $Y_i$ are the DR2 data and $Y(z_i)$ are the expressions from Eqs.~\eqref{distances}, i.e. $Y(z_i)= \{D_M(z)/r_d,
\ D_H(z)/r_d, \ D_V(z)/r_d\}$.  Furthermore, $\textbf{C}_\text{B}$ is the covariance matrix\footnote{ The full 13x13 covariance matrix can be found in the \emph{Cobaya} GitHub repository \url{https://github.com/CobayaSampler/bao_data/blob/master/desi_bao_dr2/desi_gaussian_bao_ALL_GCcomb_cov.txt}.}.

\end{enumerate}

\begin{table}
\footnotesize
\centering
\setlength{\tabcolsep}{.6em}
\renewcommand{\arraystretch}{1.2}
\begin{tabular}{lcccc}
\hline
Tracer     & $z_{eff}$ & $D_M/r_d$ & $D_H/r_d$ & $D_V/r_d$ \\
\hline
BGS & $0.295$ & $-$ & $-$ & $7.942\pm 0.075$  \\
LRG1 & $0.510$ & $13.588\pm 0.167$ & $21.863\pm 0.425$ & $-$  \\
LRG2 & $0.706$ & $17.351\pm 0.177$ & $19.455\pm 0.330$ & $-$  \\
LRG3+ELG1 & $0.934$ & $21.576\pm 0.152$ & $17.641\pm 0.193$ & $-$ \\
ELG2 & $1.321$ & $27.601\pm 0.318$ & $14.176\pm 0.221$ & $-$  \\
QSO & $1.484$ & $30.512\pm 0.760$ & $12.817\pm 0.516$ & $-$  \\
Lya QSO & $2.330$ & $38.988\pm 0.531$ & $8.632\pm 0.101$ & $-$ \\
\hline
\end{tabular}
\caption{The DR2 DESI data points with associated errors for bright galaxy survey (BGS), luminous red galaxies (LRG),
emission line galaxies (ELG), quasars (QSO), Lyman-$\alpha$ forest quasars (Lya QSO) and a combination of LRG+ELG \citep{2025arXiv250314738D}.}
\label{tab:DESIBAO}
\end{table}

\section{Constraints from gamma-ray burst correlations}\label{sec3}

GRBs are powerful extra-galactic transients prompting investigation up to $z\simeq 9$ by means of various correlations that combine different observables.

Adding GRB data is particularly helpful not only in extending constraints at higher redshifts, but also in achieving tighter constraints on $\eta_0$.
This has been already shown in \citet{2017JCAP...09..039H} where the use of a sample of $147$ GRBs -- combined with measurements of strong gravitational lensing (SGL), based on the singular isothermal sphere (SIS) or power-law index (PLAW) models, and SNe Ia -- certified the  decrease of the errors on $\eta_0$ at $2$-$\sigma$. More quantitatively, for the parametrization $\eta(z)=1+\eta_0z$, provides $\eta_0=0.15\pm 0.13$ when considering SGL (SIS)+SNe Ia+GRBs \citep{2017JCAP...09..039H} and $\eta_0=-0.005^{+0.800}_{-0.100}$ when considering only SGL (SIS)+SNe Ia  \citep{2016ApJ...822...74L}.

Here, we test the use of GRBs in improving the constraints for three CDD parametrizations. In particular, we focus on the prompt emission $E_p-E_{iso}$ (or \emph{Amati}) correlation and the prompt-afterglow emission $L_0-E_p-T$ (or \emph{Combo}) correlation and override the \emph{circularity} problem \citep{2021Galax...9...77L} using the interpolations in Eqs.~\eqref{bezier} and \eqref{da}.
\begin{itemize}
    \item [-] {\bf $E_p-E_{iso}$ correlation.} It is the most used correlation that involves only prompt emission observables \citep{2002A&A...390...81A, 2006MNRAS.372..233A}. It combines the rest-frame peak energy $E_p$ (in keV units) and the isotropic radiated energy of the bursts $E_{iso}$ (in erg units) as follows
    \begin{equation}\label{eiso0}
    \log \left(\frac{E_p}{\text{keV}}\right) = a \left[\log \left(\frac{E_{iso}}{\text{erg}}\right) - 52\right] + b,
    \end{equation}
    where $a$ and $b$ are the slope and the intercept, respectively. When CDD violations are considered through the $\eta(z)$ parametrizations, $E_{iso}$ can be written as
    \begin{equation}\label{eiso}
        E_{iso}(z) = 4\pi D_2^2(z)\eta^2(z)(1+z)^3S_b,
    \end{equation}
    with the bolometric fluence $S_b$ in erg/cm$^2$/s units.
    The ln-likelihood for this correlation, with a data set of $N_A = 118$ sources, considering that in this case from Eq. \eqref{chi2} $X_i-X(z_i)=\log E_{iso,i}-\log E_{iso}(z_i)$, is given by \citep{2021JCAP...09..042K}
    \begin{equation}
        \ln\mathcal{L}_A\propto -\frac{1}{2}\sum^{N_A}_{i=1}\frac{\left[\log E_{iso,i}-\log E_{iso}(z_i)\right]^2}{\sigma^2_{\log S_{b,i}}+a^{-2}\sigma^2_{\log E_{p,i}}+a^{-2}\sigma^2},
    \end{equation}
    where $\log E_{iso,i}$ and $\log E_{iso}(z_i)$ are defined by Eqs.~\eqref{eiso0} and \eqref{eiso}, respectively, and $\sigma$ is the intrinsic dispersion of the correlation \citep{DAgostini:2005mth}.
    \item [-] {\bf $L_0-E_p-T$ correlation.} It considers both quantities from the prompt and X-ray afterglow emissions \citep{2015A&A...582A.115I} and it is obtained by combining the $E_p-E_{iso}$ and the $E^X_{iso}-E_{iso}-E_p$ correlations \citep{2006MNRAS.372..233A, 2012MNRAS.425.1199B}
    \begin{equation}\label{l00}
        \log \left(\frac{L_0}{\text{erg/s}}\right) = a\log \left(\frac{E_p}{\text{keV}}\right) - \log \left(\frac{T}{s}\right) + b,
    \end{equation}
    where, as for the $E_p-E_{iso}$ correlation, $a$ is the slope while $b$ is the intercept of the correlation, $T$ (in units of s) represents the effective rest-frame duration of plateau phase of the X-ray afterglow. The plateau luminosity $L_0$ (in units of erg/s), when we consider a violation of the CDD relation, is defined in terms of the flux of the plateau $F_0$ as
    \begin{equation}\label{l0}
        L_0(z) = 4\pi D_2^2(z)\eta^2(z)(1+z)^4F_0.
    \end{equation}
    The ln-likelihood for this correlation, considering that in this case from Eq. \eqref{chi2} $X_i-X(z_i)=\log L_{0,i}-\log L_0(z_i)$, is \citep{2021JCAP...09..042K}
    \begin{equation}
        \ln\mathcal{L}_C\propto -\frac{1}{2}\sum^{N_C}_{i=1}\frac{\left[\log L_{0,i}-\log L_0(z_i)\right]^2}{\sigma^2_{\log L_{0,i}}\!+a^2\sigma^2_{\log E_{p,i}}\!+\sigma^2_{\log T_i}\!+\sigma^2},
    \end{equation}
    where $N_C = 173$ and $\log L_{0,i}$ and $\log L_0(z_i)$ are defined by Eqs.~\eqref{l00} and \eqref{l0}, respectively. Again, $\sigma$ is the intrinsic dispersion of the correlation \citep{DAgostini:2005mth}.
\end{itemize}

\section{Numerical results}\label{sec4}

We perform two analyses that differ, specifically, for the involved GRB correlations. The GRB data sets used in these analyses are taken from Ref.~\cite{2021JCAP...09..042K}
\begin{enumerate}
    \item [-] {\bf Analysis A} considers the A118 sample of GRB fulfilling the $E_p-E_{iso}$ correlation \citep{2021JCAP...09..042K}, the updated OHD sample, the $N_{GC}$ angular diameter distances as inferred from galaxy clusters, the Pantheon sample and the DR2 from DESI.
    \item [-] {\bf Analysis C} considers the same samples as Analysis A, but replaces the A118 data set with the C173 sample of the $L_0-E_p-T$ correlation \citep{2021JCAP...09..042K}.
\end{enumerate}

In all analyses the cosmological constraints are obtained by maximizing the total ln-likelihood which is the sum of the ln-likelihood of each used sample
\begin{align*}
    {\text{\bf Analysis A}}:  \ln\mathcal{L} = \ln\mathcal{L}_{A}+\ln\mathcal{L}_{\chi},\\
    {\text{\bf Analysis C}}:  \ln\mathcal{L} = \ln\mathcal{L}_{C}+\ln\mathcal{L}_{\chi},
\end{align*}
where $\ln\mathcal{L}_{\chi}=\ln\mathcal{L}_{O}+\ln\mathcal{L}_{GC}+\ln\mathcal{L}_{SN}+\ln\mathcal{L}_D$.

\begin{figure*}
\centering
{\hfill
\includegraphics[width=0.47\hsize,clip]{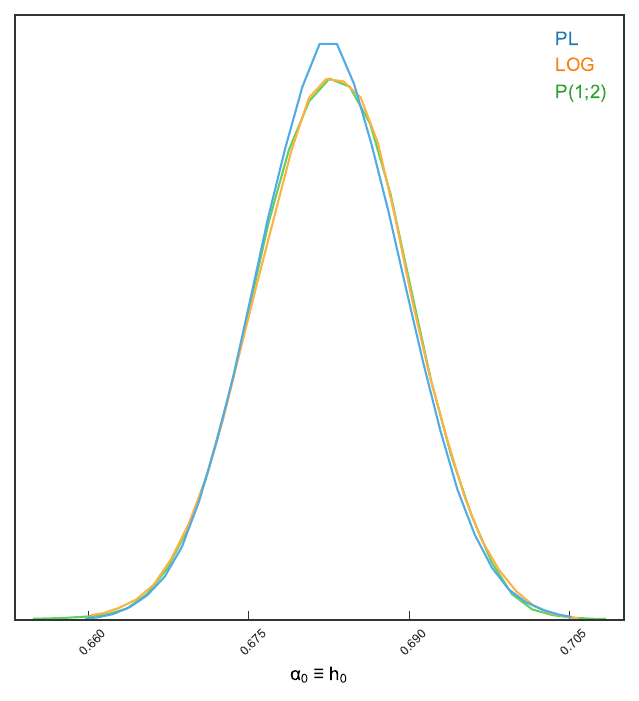}
\hfill
\includegraphics[width=0.465\hsize,clip]{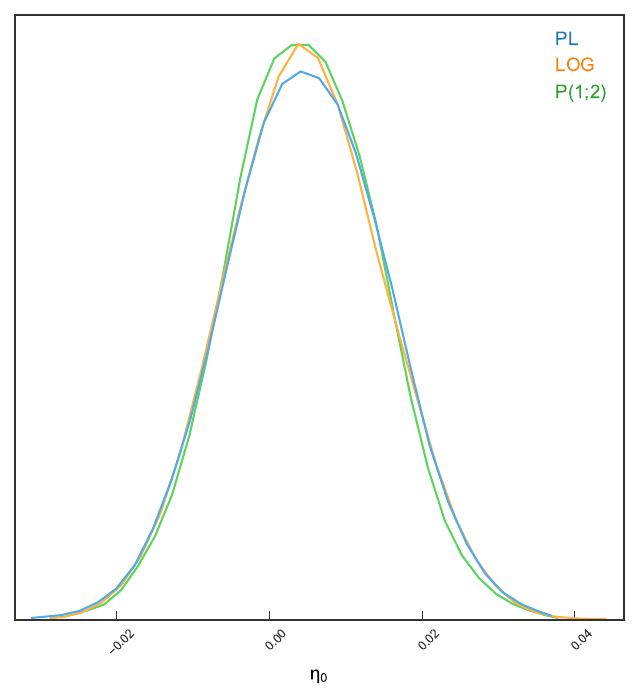}
\hfill}\\
\caption{1D posterior distributions for the parameters $\alpha_0\equiv h_0$ (left plot), and $\eta_0$ (right plot). Both plots are obtained considering the combination of data sets in \emph{Analysis A}. The colors are related to the parametrizations of $\eta(z)$ used in our computations: blue for the power-law parametrization, orange for the LOG model and green for the $(1;2)$ Pad\'e series.}
 \label{fig:Amati}
\end{figure*}

\begin{figure*}
\centering
{\hfill
\includegraphics[width=0.47\hsize,clip]{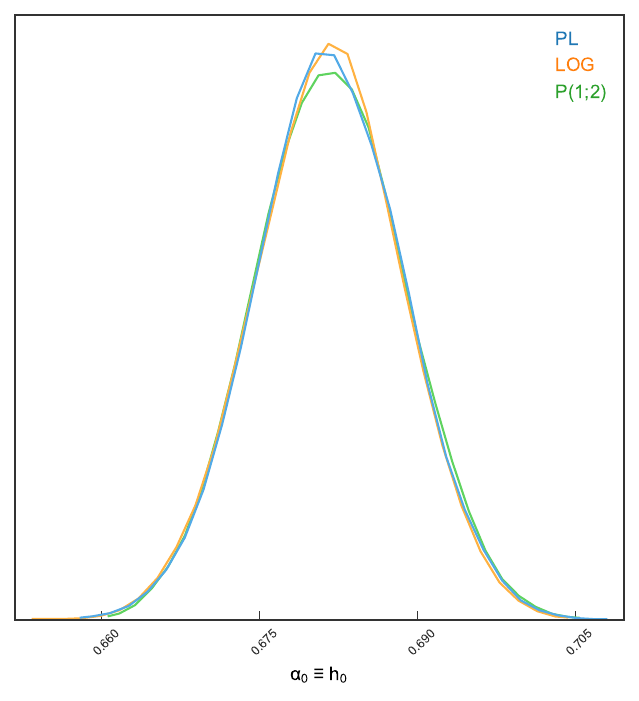}
\hfill
\includegraphics[width=0.465\hsize,clip]{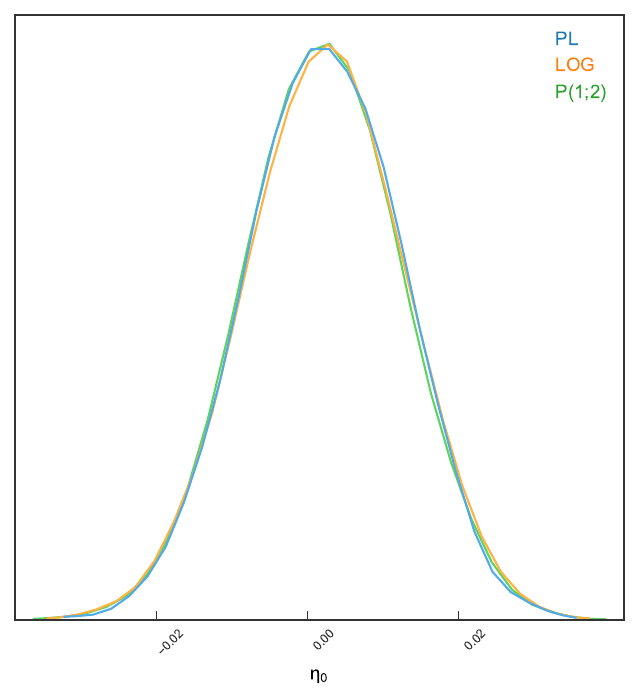}
\hfill}\\
\caption{Same 1D posteriors and color code of Fig.~\ref{fig:Amati} here obtained by considering the combination of data sets in \emph{Analysis C}.}
 \label{fig:Combo}
\end{figure*}

In the following we discuss the results listed in Tabs.~\ref{tab:bfAmati}-\ref{tab:bfCombo} and portrayed in  Figs.~\ref{fig:Amati}-\ref{fig:Combo} by the one-dimensional (1D) posterior distributions for $h_0$ and $\eta_0$ obtained using the python-based code pygtc \citep{Bocquet2016}.

\begin{table*}
\scriptsize
\centering
\setlength{\tabcolsep}{2.em}
\renewcommand{\arraystretch}{1.5}
\begin{tabular}{ccccccc}
\hline
 $a$ & $b$ & $\sigma$ & $\alpha_0\equiv h_0$ & $\alpha_1$ & $\alpha_2$ & $\eta_0$ \\
\hline
\multicolumn{7}{c}{PL}\\
\cline{1-7}
$0.854^{+0.099(0.172)}_{-0.087(0.136)}$ & $1.618^{+0.131(0.203)}_{-0.129(0.227)}$ & $0.327^{+0.066(0.119)}_{-0.046(0.074)}$ & $0.683^{+0.010(0.017)}_{-0.011(0.016)}$ & $1.054^{+0.027(0.046)}_{-0.025(0.043)}$ & $2.002^{+0.024(0.041)}_{-0.030(0.046)}$ & $0.003^{+0.016(0.027)}_{-0.014(0.024)}$\\
\hline
\multicolumn{7}{c}{LOG}\\
\cline{1-7}
$0.856^{+0.090(0.163)}_{-0.085(0.133)}$ & $1.621^{+0.124(0.190)}_{-0.125(0.235)}$ & $0.337^{+0.049(0.103)}_{-0.056(0.080)}$ & $0.682^{+0.012(0.019)}_{-0.009(0.016)}$ & $1.060^{+0.022(0.038)}_{-0.032(0.048)}$ & $1.996^{+0.029(0.046)}_{-0.023(0.039)}$ & $0.004^{+0.017(0.025)}_{-0.015(0.023)}$\\
\hline
\multicolumn{7}{c}{P(1,2)}\\
\cline{1-7}
$0.856^{+0.099(0.178)}_{-0.087(0.135)}$ & $1.625^{+0.121(0.189)}_{-0.144(0.245)}$ & $0.332^{+0.062(0.111)}_{-0.052(0.077)}$ & $0.683^{+0.010(0.017)}_{-0.011(0.018)}$ & $1.052^{+0.027(0.046)}_{-0.023(0.040)}$ & $2.000^{+0.026(0.043)}_{-0.028(0.044)}$ & $0.005^{+0.014(0.025)}_{-0.015(0.025)}$\\
\hline
\end{tabular}
\caption{\emph{Analysis A} best fits, $1$-$\sigma$ ($2$-$\sigma$) errors of GRB correlation parameters, B\'ezier coefficients, and $\eta_0$, respectively.}
\label{tab:bfAmati}
\end{table*}

\subsection{Analysis A}

We discuss the outcomes of our MCMC analysis focusing on the $\alpha_0$ B\'ezier coefficient, which coincides with the reduced Hubble constant $h_0$, and $\eta_0$ obtained from the three parametrizations discussed in this work: power-law, LOG model and Pad\'e series of order $(1;2)$ or P(1;2).
\begin{enumerate}
    \item [-] {\bf Power-law.} We find that $\eta_0$ is compatible with zero at $1$-$\sigma$. Regarding the Hubble tension, our $h_0$ agrees with Planck measurement $h^P_0=0.674\pm 0.005$ \citep{2020A&A...641A...6P} at $1$-$\sigma$, while we found no compatibility with $h^S_0=0.730\pm 0.010$ derived by \citet{2022ApJ...934L...7R}.
    \item [-] {\bf LOG model.} In this case $\eta_0$ is consistent with zero at $1$-$\sigma$, hence for this model no deviation from the CDD relation subsists. Regarding the $h_0$ tension, we assess that our $h_0$ agrees only with the value of the \citet{2020A&A...641A...6P} at $1$-$\sigma$, while no compatibility is found with that of \citet{2022ApJ...934L...7R}.
    \item [-] {\bf P(1;2) series.} Also here $\eta_0\approx0$ at $1$-$\sigma$, meaning no deviation from the CDD relation. Again, our $h_0$ agrees with the value of the \citet{2020A&A...641A...6P} at $1$-$\sigma$, while no compatibility exists with the value found by \cite{2022ApJ...934L...7R}.
\end{enumerate}

\begin{table*}
\scriptsize
\centering
\setlength{\tabcolsep}{2.em}
\renewcommand{\arraystretch}{1.5}
\begin{tabular}{ccccccc}
\hline
 $a$ & $b$ & $\sigma$ & $\alpha_0\equiv h_0$ & $\alpha_1$ & $\alpha_2$ & $\eta_0$ \\
\hline
\multicolumn{7}{c}{PL}\\
\cline{1-7}
$0.821^{+0.105(0.172)}_{-0.080(0.159)}$ & $49.702^{+0.215(0.422)}_{-0.269(0.456)}$ & $0.368^{+0.044(0.072)}_{-0.036(0.060)}$ & $0.681^{+0.011(0.018)}_{-0.009(0.015)}$ & $1.057^{+0.022(0.039)}_{-0.029(0.046)}$ & $1.998^{+0.026(0.042)}_{-0.026(0.043)}$ & $0.001^{+0.016(0.025)}_{-0.014(0.025)}$\\
\hline
\multicolumn{7}{c}{LOG}\\
\cline{1-7}
$0.844^{+0.081(0.154)}_{-0.117(0.190)}$ & $49.655^{+0.300(0.484)}_{-0.220(0.412)}$ & $0.367^{+0.042(0.069)}_{-0.034(0.056)}$ & $0.682^{+0.010(0.017)}_{-0.011(0.017)}$ & $1.056^{+0.024(0.041)}_{-0.027(0.046)}$ & $1.999^{+0.025(0.042)}_{-0.026(0.044)}$ & $0.004^{+0.015(0.024)}_{-0.017(0.027)}$\\
\hline
\multicolumn{7}{c}{P(1,2)}\\
\cline{1-7}
$0.828^{+0.092(0.167)}_{-0.095(0.171)}$ & $49.671^{+0.262(0.462)}_{-0.221(0.421)}$ & $0.372^{+0.038(0.067)}_{-0.040(0.061)}$ & $0.681^{+0.011(0.018)}_{-0.010(0.016)}$ & $1.053^{+0.027(0.043)}_{-0.025(0.044)}$ & $1.999^{+0.026(0.045)}_{-0.028(0.041)}$ & $0.002^{+0.015(0.025)}_{-0.015(0.025)}$\\
\hline
\end{tabular}
\caption{\emph{Analysis C} best fits, $1$-$\sigma$ ($2$-$\sigma$) errors of GRB correlation parameters, B\'ezier coefficients, and $\eta_0$, respectively.}
\label{tab:bfCombo}
\end{table*}

\subsection{Analysis C}

Now, moving to the results drawn from using the $L_0-E_p-T$ correlation, as for \emph{Analysis A}, we discuss the constraints obtained on $\alpha_0\equiv h_0$ and $\eta_0$ for the three parametrizations of $\eta(z)$ displayed in Eqs.~\eqref{stepintermedio}-\eqref{P12}.
\begin{itemize}
    \item [-] {\bf Power-law.} We find that $\eta_0$ is compatible with zero within $1$-$\sigma$. Regarding the Hubble tension, our $h_0$ is compatible with the value of \citet{2020A&A...641A...6P} at $1$-$\sigma$, while the measurement obtained from SNe~Ia \citep{2022ApJ...934L...7R} is incompatible.
    \item [-] {\bf LOG model.} This scenario excludes CDD violations since $\eta_0$ at $1$-$\sigma$ is totally compatible with zero. Comparing $h_0$ with the measurements from Planck and Riess, we find an agreement at $1$-$\sigma$ with the former and no compatibility with the latter.
    \item [-] {\bf P(1;2) series.} Here, we find that $\eta_0$ agrees at $1$-$\sigma$ with zero showing also in this case no deviations from the CDD relation. Finally, our results for $h_0$ agree at $1$-$\sigma$ with $h_0^P=0.674\pm 0.005$ \citep{2020A&A...641A...6P} and not at all with $h_0^S=0.730\pm 0.010$ \citep{2022ApJ...934L...7R}.
\end{itemize}

\section{Final outlooks}\label{sec5}

In this paper,  we investigated possible violations of the CDD relation up to $z\simeq 9$ through the inclusion of two GRB correlations, the $E_p-E_{iso}$ and the $L_0-E_p-T$ correlations, together with low-redshift catalogs such as OHD, galaxy clusters, Pantheon SNe~Ia, and BAO from the DR2 of DESI. In doing so, we considered that, if a violation of Etherington's reciprocity law would exist, then the ratio between the luminosity and angular diameter distances would take the form $d_L/[d_A(1+z)^2]=\eta(z)$.

A violation of this relation would imply the search for possible systematics in the determination of the distances or the need to investigate new physics related to, e.g. a violation of the conservation of  photon number \citep{2004PhRvD..69j1305B}.

Hence, we proposed three parametrizations of $\eta(z)$ mimicking possible departures from the relation. First, we considered a power-law violation $(1+z)^{\eta_0}$, also used to investigate possible violation of the \emph{cosmic transparency} \citep{2009JCAP...06..012A, 2018PhRvD..97b3538H, 2025JCAP...01..088J}. Then, assuming that $|\eta_0|\ll 1$, we expanded the power-law parametrization leading to logarithmic correction with a slower redshift evolution. Finally, we used Pad\'e rational polynomials of order $(1;2)$ in order to heal possible convergence problems.

To test the aforementioned parametrizations, we followed a model-independent approach by interpolating the $34$ data points from the OHD catalog through a second order B\'ezier parametric curve \citep{2019MNRAS.486L..46A}. Then, assuming a flat universe scenario, the so-reconstructed Hubble rate $H(z)$ is employed to interpolate the observables from the other catalogs employed in this work, including GRB data from the $E_p-E_{iso}$ and the $L_0-E_p-T$ correlations, for which the interpolation overcomes the \emph{circularity} problem \citep{2021Galax...9...77L}.

Then, we run two MCMC fits based on the Metropolis-Hastings algorithm \citep{1953JChPh..21.1087M, 1970Bimka..57...97H} employing catalogs from OHD, galaxy clusters, SNe Ia, and BAO, including GRBs from the $E_p-E_{iso}$ correlation in \emph{Analysis A}, or from the $L_0-E_p-T$ correlation in \emph{Analysis C}.

We found out that there is no evidence for violations of the CDD relation in both \emph{Analysis A} and \emph{Analysis C} since our results favor $\eta_0\approx0$ within $1$-$\sigma$. Afterwards, focusing on the Hubble tension, we compared our results on the (reduced) Hubble rate $h_0$ with $h_0^P=0.674\pm 0.005$ inferred by the \citet{2020A&A...641A...6P} and with $h_0^S=0.730\pm 0.010$ inferred by \citet{2022ApJ...934L...7R}.
Our outcomes, in both \emph{Analysis A} and \emph{Analysis C}, are in agreement only with the \citet{2020A&A...641A...6P} value at $1$-$\sigma$ while no compatibility has been found when the $h^S_0$ from SNe Ia was accounted for.

Future works will focus on investigating more deeply this last point on the Hubble tension using instead of GRBs other high-redshift probes, such as standard sirens, the gravitational wave analog of standard candles arising from compact object mergings. Specifically, with the advent of the Einstein Telescope (ET), it would be possible to obtain $\sim 1000$ events after ten years of operation \citep{2010CQGra..27u5006S}. This would provide a catalog larger than that emerging from GRBs.

\section*{Acknowledgements}

ACA and SC acknowledge the support of the Istituto Nazionale di Fisica Nucleare (INFN) Sezione di Napoli, Iniziativa Specifica QGSKY. SC thanks the Gruppo Nazionale di Fisica Matematica (GNFM) of Istituto Nazionale di Alta Matematica (INDAM) for the support.
OL acknowledges support by the  Fondazione  ICSC, Spoke 3 Astrophysics and Cosmos Observations. National Recovery and Resilience Plan (Piano Nazionale di Ripresa e Resilienza, PNRR) Project ID $CN00000013$ ``Italian Research Center on  High-Performance Computing, Big Data and Quantum Computing" funded by MUR Missione 4 Componente 2 Investimento 1.4: Potenziamento strutture di ricerca e creazione di ``campioni nazionali di R\&S (M4C2-19)" - Next Generation EU (NGEU). MM acknowledges support from the project OASIS, ``PNRR Bando a Cascata da INAF M4C2 - INV. 1.4''.
MM acknowledges the support of the European Union - NextGenerationEU, Mission 4, Component 2, under the Italian Ministry of University and Research (MUR) - Strengthening research structures and creation of "national R\&D champions" on some Key Enabling Technologies - grant CN00000033 - NBFC - CUPJ13C23000490006.
This paper is based upon work from COST Action CA21136 -- Addressing observational tensions in cosmology with systematics and fundamental physics (CosmoVerse), supported by COST (European Cooperation in Science and Technology).


\bibliographystyle{elsarticle-harv}






\end{document}